\documentclass[superscriptaddress,prx,longbibliography,nofootinbib,notitlepage,10pt]{revtex4-1}
\pdfoutput=1
\usepackage{graphicx}
\usepackage{amsmath}
\usepackage{amssymb}
\usepackage{amsthm}
\usepackage{comment}
\usepackage{cprotect}
\usepackage{placeins}
\usepackage[caption=false]{subfig}
\usepackage[colorlinks]{hyperref}
\usepackage[all]{hypcap}
\usepackage{tikz}
\usepackage{verbatim}
\usepackage{subfig}
\usetikzlibrary{arrows}
\usepackage{units}
\usepackage{soul}
\usepackage{textcomp}
\usepackage{fancyvrb}

\usepackage{listings}
\usepackage{color}
\usepackage[utf8]{inputenc}
\definecolor{codegreen}{rgb}{0,0.6,0}
\definecolor{codegray}{rgb}{0.5,0.5,0.5}
\definecolor{codepurple}{rgb}{0.58,0,0.82}
\definecolor{backcolour}{rgb}{0.95,0.95,0.92}

\lstdefinestyle{mystyle}{
  backgroundcolor=\color{backcolour},   commentstyle=\color{codegreen},
  keywordstyle=\color{magenta},
  numberstyle=\tiny\color{codegray},
  stringstyle=\color{codepurple},
  basicstyle=\footnotesize,
  breakatwhitespace=false,         
  breaklines=true,                 
  captionpos=b,                    
  keepspaces=true,                 
  numbers=left,                    
  numbersep=5pt,                  
  showspaces=false,                
  showstringspaces=false,
  showtabs=false,                  
  tabsize=2
}

\newcommand{\eq}[1]{Eq.~\hyperref[eq:#1]{(\ref*{eq:#1})}}
\renewcommand{\sec}[1]{\hyperref[sec:#1]{Section~\ref*{sec:#1}}}
\DeclareRobustCommand{\app}[1]{\hyperref[app:#1]{Appendix~\ref*{app:#1}}}
\newcommand{\tab}[1]{\hyperref[tab:#1]{Table~\ref*{tab:#1}}}
\newcommand{\fig}[1]{\hyperref[fig:#1]{Figure~\ref*{fig:#1}}}
\newcommand{\figa}[2]{\hyperref[fig:#1]{Figure~\ref*{fig:#1}#2}}
\newcommand{\figx}[2]{\hyperref[fig:#1]{Figure~\ref*{fig:#1}(#2)}}
\newcommand{\thm}[1]{\hyperref[thm:#1]{Theorem~\ref*{thm:#1}}}
\newcommand{\lem}[1]{\hyperref[lem:#1]{Lemma~\ref*{lem:#1}}}
\newcommand{\cor}[1]{\hyperref[cor:#1]{Corollary~\ref*{cor:#1}}}
\newcommand{\defn}[1]{\hyperref[def:#1]{Definition~\ref*{def:#1}}}
\newcommand{\alg}[1]{\hyperref[alg:#1]{Algorithm~\ref*{alg:#1}}}

\def\avg#1{\mathinner{\langle{#1}\rangle}}
\def\bra#1{\mathinner{\langle{#1}|}}
\def\ket#1{\mathinner{|{#1}\rangle}}

\input{Qcircuit}

\begin{document}

\title{
OpenFermion: 
The Electronic Structure Package for Quantum Computers}

\date{\today}

\author{Jarrod R.\ McClean}
\email[Corresponding author: ]{jarrodmcc@gmail.com}
\affiliation{Google Inc., Venice, CA 90291}
\author{Kevin J. Sung}
\affiliation{Department of Electrical Engineering and Computer Science, University of Michigan, Ann Arbor, MI 48109}
\author{Ian D.\ Kivlichan}
\affiliation{Google Inc., Venice, CA 90291}
\affiliation{Department of Physics, Harvard University, Cambridge, MA 02138}

\author{Yudong Cao}
\affiliation{Department of Chemistry and Chemical Biology, Harvard University, Cambridge, MA 02138}
\affiliation{Zapata Computing Inc., Cambridge 02138}
\author{Chengyu Dai}
\affiliation{Department of Physics, University of Michigan, Ann Arbor, MI 48109}
\author{E.\ Schuyler Fried}
\affiliation{Department of Chemistry and Chemical Biology, Harvard University, Cambridge, MA 02138}
\affiliation{Rigetti Computing, Berkeley CA 94710}
\author{Craig Gidney}
\affiliation{Google Inc., Santa Barbara, CA 93117}
\author{Brendan Gimby}
\affiliation{Department of Electrical Engineering and Computer Science, University of Michigan, Ann Arbor, MI 48109}
\author{Pranav Gokhale}
\affiliation{Department of Computer Science, University of Chicago, Chicago, IL 60637}
\author{Thomas H\"{a}ner}
\affiliation{Theoretische Physik, ETH Zurich, 8093 Zurich, Switzerland}
\author{Tarini Hardikar}
\affiliation{Department of Physics, Dartmouth College, Hanover, NH 03755}
\author{Vojt\v ech Havl\' i\v cek}
\affiliation{Department of Computer Science, Oxford University, Oxford OX1 3QD, United Kingdom}
\author{Oscar Higgott}
\affiliation{Department of Physics and Astronomy, University College London, Gower Street, London WC1E 6BT, United Kingdom}
\author{Cupjin Huang}
\affiliation{Department of Electrical Engineering and Computer Science, University of Michigan, Ann Arbor, MI 48109}
\author{Josh Izaac}
\affiliation{Xanadu, 372 Richmond St W, Toronto, M5V 1X6, Canada}
\author{Zhang Jiang}
\affiliation{Google Inc., Venice, CA 90291}
\affiliation{QuAIL, NASA Ames Research Center, Moffett Field, CA 94035}
\author{Xinle Liu}
\affiliation{Google Inc., Mountain View, CA 94043}
\author{Sam McArdle}
\affiliation{Department of Materials, University of Oxford, Parks Road, Oxford OX1 3PH, United Kingdom}
\author{Matthew Neeley}
\affiliation{Google Inc., Santa Barbara, CA 93117}
\author{Thomas O'Brien}
\affiliation{Instituut-Lorentz, Universiteit Leiden, 2300 RA Leiden, The Netherlands}
\author{Bryan O'Gorman}
\affiliation{BQIC, University of California, Berkeley, CA 94720}
\affiliation{QuAIL, NASA Ames Research Center, Moffett Field, CA 94035}
\author{Isil Ozfidan}
\affiliation{D-Wave Systems Inc., Burnaby, BC}
\author{Maxwell D.\ Radin}
\affiliation{Materials Department, University of California Santa Barbara, Santa Barbara, CA 93106}
\author{Jhonathan Romero}
\affiliation{Department of Chemistry and Chemical Biology, Harvard University, Cambridge, MA 02138}
\author{Nicholas Rubin}
\affiliation{Rigetti Computing, Berkeley CA 94710}
\author{Nicolas P.\ D.\ Sawaya}
\affiliation{Department of Chemistry and Chemical Biology, Harvard University, Cambridge, MA 02138}
\author{Kanav Setia}
\affiliation{Department of Physics, Dartmouth College, Hanover, NH 03755}
\author{Sukin Sim}
\affiliation{Department of Chemistry and Chemical Biology, Harvard University, Cambridge, MA 02138}
\author{Damian S.\ Steiger}
\affiliation{Google Inc., Venice, CA 90291}
\affiliation{Theoretische Physik, ETH Zurich, 8093 Zurich, Switzerland}
\author{Mark Steudtner}
\affiliation{Instituut-Lorentz, Universiteit Leiden, 2300 RA Leiden, The Netherlands}
\affiliation{QuTech, Delft University of Technology, Lorentzweg 1, 2628 CJ Delft, The Netherlands}
\author{Qiming Sun}
\affiliation{Division of Chemistry and Chemical Engineering, California Institute of Technology, Pasadena, CA 91125}
\author{Wei Sun}
\affiliation{Google Inc., Mountain View, CA 94043}
\author{Daochen Wang}
\affiliation{Joint Center for Quantum Information and Computer Science, University of Maryland, College Park, MD 20742}
\author{Fang Zhang}
\affiliation{Department of Electrical Engineering and Computer Science, University of Michigan, Ann Arbor, MI 48109}

\author{Ryan Babbush}
\email[Corresponding author: ]{ryanbabbush@gmail.com}
\affiliation{Google Inc., Venice, CA 90291}

\begin{abstract}
Quantum simulation of chemistry and materials is predicted to be an important application for both near-term and fault-tolerant quantum devices. However, at present, developing and studying algorithms for these problems can be difficult due to the prohibitive amount of domain knowledge required in both the area of chemistry and quantum algorithms. To help bridge this gap and open the field to more researchers, we have developed the OpenFermion software package (\href{http://www.openfermion.org}{\texttt{www.openfermion.org}}). OpenFermion is an open-source software library written largely in Python under an Apache 2.0 license, aimed at enabling the simulation of fermionic and bosonic models and quantum chemistry problems on quantum hardware. Beginning with an interface to common electronic structure packages, it simplifies the translation between a molecular specification and a quantum circuit for solving or studying the electronic structure problem on a quantum computer, minimizing the amount of domain expertise required to enter the field. The package is designed to be extensible and robust, maintaining high software standards in documentation and testing. This release paper outlines the key motivations behind design choices in OpenFermion and discusses some basic OpenFermion functionality which we believe will aid the community in the development of better quantum algorithms and tools for this exciting area of research.
\end{abstract}
\maketitle

\section*{Introduction}

Recent strides in the development of hardware for quantum computing demand comparable developments in the applications
and software these devices will run.  Since the inception of quantum computation, a number of promising application
areas have been identified, ranging from factoring~\cite{Shor1997} and solutions of linear equations~\cite{Harrow:2009}
to simulation of complex quantum materials.  However, while the theory has been well developed for many of these problems, the challenge of compiling efficient algorithms
for these devices down to hardware realizable gates remains a formidable one.  While this problem is difficult to tackle in full generality,
significant progress can be made in particular areas.  Here, we focus on what was perhaps the original application for quantum
devices, quantum simulation.

Beginning with Feynman in 1982~\cite{Feynman1982}, it was proposed that highly controllable quantum devices, 
later to become known as quantum computers,
would be especially good at the simulation of other quantum systems.  This notion was later formalized to show how in particular
instances, we expect an exponential speedup in the solution of the Schr\"odinger equation for 
chemical systems~\cite{Lloyd1996,Abrams1997,Abrams1999,Ortiz2001,Somma2002,Aspuru-Guzik2005}. This opens
the possibility of understanding and designing new materials, drugs, and catalysts that were previously untenable. There has also been substantial work applying these methods to other fermionic systems such as the Hubbard model \cite{Jiang2017,Wecker2015b,Kivlichan2018} and Sachdev-Ye-Kitaev model \cite{Garcia-Alvarez2017,BabbushSYK}.  Since
the initial work in this area, there has been great progress developing new algorithms \cite{Kassal2008,Ward2009,Whitfield2010,Toloui2013,Hastings2015,BabbushSparse2,BabbushSparse1,BabbushAQChem,Kivlichan2016,Sugisaki2016,BabbushLow,Motzoi2017,Kivlichan2017,Berry2018,BabbushContinuum,Berry2019}, tighter bounds and better implementation strategies \cite{Veis2010,Veis2014,McClean2014,Wecker2014,Poulin2014,BabbushTrotter,Kivlichan2017,Berry2018,Rubin2018}, more desirable Hamiltonian representations \cite{Seeley2012,Whitfield2013b,Tranter2015,Moll2015,Whitfield2016,Barkoutsos2017,Havlicek2017,Bravyi2017,Zhu2017,Setia2017,Steudtner2017,Motta2018,Jiang2018}, fault-tolerant resource estimates and layouts \cite{Jones2012,Reiher2017,BabbushSpectra}, and proof-of-concept experimental
demonstrations~\cite{Lanyon2010,Li2011,Wang2014,Santagati2016}. Moreover, with mounting experimental evidence, variational and hybrid-quantum classical
algorithms~\cite{Peruzzo2013,Yung2013,McClean2015,Shen2015,Wecker2015a,sawaya2016error,McClean2017a,OMalley2016,Colless:2017,Kandala2017,Romero2017,Hempel2018} for these systems have been identified as particularly promising approaches when one has limited resources; some speculate these quantum algorithms may even solve classically intractable instances without error-correction.

However, despite immense progress, much work is left to be done in optimizing algorithms in this area for both near-term and far-future devices. Already this field has seen instances where the difference between naive bounds for algorithms and expected numerics for real systems can differ by many orders of magnitude~\cite{Wecker2014,Hastings2015,Poulin2014,BabbushTrotter,Reiher2017}. Unfortunately, developing algorithms in this area can require a prohibitive amount of domain expertise. For example, quantum algorithms experts may find the chemistry literature rife with jargon and unknown approximations while chemists find themselves unfamiliar with the concepts used in quantum information. As has been seen though, both have a crucial role to play in developing algorithms for these emerging devices.

For this reason, we introduce the OpenFermion package, designed to bridge the gap between different domain areas and facilitate
the development of explicit quantum simulation algorithms for quantum chemistry.  The goal of this project is to enable both
quantum algorithm developers and quantum chemists to make contributions to this developing field, minimizing the amount of domain
knowledge required to get started, and aiding those with knowledge to perform their work more quickly and easily.  OpenFermion is an
open-source Apache 2.0 licensed software package to encourage community adoption and contribution.  It has a modular design
and is maintained with strict documentation and testing standards to ensure robust code and project longevity.  Moreover, to maximize usefulness within the field, every effort has been made to design OpenFermion as a modular library which is agnostic with respect to quantum programming language frameworks.  Through its plugin system, OpenFermion is able to interface with, and benefit from, any of the frameworks being developed for both more abstract quantum software and hardware specific compilation~\cite{Steiger2016,Cross:2017,Smith:2016,Green:2013,Selinger2004,liq1,Valiron:2015,heckey2015compiler,javadiabhari2014scaffcc,Fried:2017,Killoran2018}.

This technical document introduces the first release of the OpenFermion package and proceeds in the following way.  We begin
by mapping the standard workflow of a researcher looking to implement an electronic structure problem on a quantum computer,
giving examples along this path of how OpenFermion aids the researcher in making this process as painless as possible.
We aim to make this exposition accessible to all interested readers, so some background in the problem is provided as well.
The discussion then shifts to the core and derived data structures the package is designed around.  Following this, some example
applications from real research projects are discussed, demonstrating real-world examples of how OpenFermion can streamline
the research process in this area.  Finally, we finish with a brief discussion of the open source philosophy of the project
and plans for future developments.

\section{Quantum workflow pipeline}

In this section we step through one of the primary problems in entering quantum chemistry problems to a quantum computer.
Specifically, the workflow for translating a problem in quantum chemistry to one in quantum computing.  This process
begins by specifying the problem of interest, which is typically finding some molecular property for a particular state
of a molecule or material.  This requires one to specify the molecule of interest, usually through the positions of the atoms
and their identities.  However, this problem is also steeped in some domain specific terminology, especially with regards to
symmetries and choices of discretizations.  OpenFermion helps to remove many of these particular difficulties.  Once the 
problem has been specified, several computational intermediates must be computed, namely the bare two-electron integrals
in the chosen discretization as well as the transformation to a molecular orbital basis that may be needed for certain
correlated methods.  From there, translation to qubits may be performed by one of several mappings.  The goal of this section
is to both detail this pipeline, and show at each step how OpenFermion may be used to help perform it with ease.

\subsection{Molecule specification and input generation}
Electronic structure typically refers to the problem of determining the electronic configuration for a fixed set
of nuclear positions assuming non-relativistic energy scales.  To begin with, we show how a molecule is defined within
OpenFermion, then continue on to describe what each component represents.
\begin{lstlisting}[language=Python, upquote=True, caption=Defining a simple H$_2$ instance in OpenFermion.]
from openfermion.hamiltonians import MolecularData
geometry = [['H', [0, 0, 0]],
            ['H', [0, 0, 0.74]]]
basis = 'sto-3g'
multiplicity = 1
charge = 0
h2_molecule = MolecularData(geometry, basis, multiplicity, charge)
\end{lstlisting}
The specification of a molecular geometry implicitly assumes the Born-Oppenheimer approximation, 
which treats the nuclei as
fixed point charges, and the ground state electronic energy is a parametric function of their positions.
When the positions of the nuclei are specified, the electronic structure problem can be restated as
finding the eigenstates of the Hamiltonian operator
\begin{align}
H(\mathbf{r}; \mathbf{R}) = -\sum_i \frac{\nabla^2_{r_i}}{2} - \sum_{i, j}\frac{Z_j}{|R_j - r_i|} + \sum_{i < j} \frac{1}{|r_i - r_j|} + \sum_{i < j}\frac{Z_i Z_j}{|R_i - R_j|}
\end{align}
where we have used atomic units (i.e. $\hbar=1$), $r_i$ represent the positions of electrons, $R_i$ represent the positions of nuclei, and $Z_i$ are the charges of nuclei.  In our example from OpenFermion, we see that the nuclear positions may
be specified by a set of $x,y,z$ coordinates along with atom labels as
\begin{lstlisting}[language=Python]
geometry = [[AtomLabel1, [x_1, y_1, z_1]],
            [AtomLabel2, [x_2, y_2, z_2]],
            ...]
\end{lstlisting}
A reasonable set of nuclear positions $R$ for a given molecule can often be given from experimental data, 
empirical models, or a geometry optimization.  A geometry optimization minimizes
the lowest eigenvalue of the above Hamiltonian as a function of the nuclear position, to some local, stable optimum.
These nuclear configurations are sometimes called equilibrium structures with respect to the level of theory used
to calculate them.

As we focus on the non-relativistic Hamiltonian, there is no explicit dependence on electron spin. As a result,
the total electronic spin and one of its components (canonically chosen to be the $Z$ direction) 
form good quantum numbers for the Hamiltonian.  That is, the Hamiltonian can be made block diagonal with separate 
spin labels for each block.  Explicitly including this symmetry offers a number of computational advantages, including
smaller spaces to explore, as well as the ability to access excited states that are the lowest energy state within
a particular spin manifold using ground state methods \cite{Helgaker2002}.  In particular, we parameterize these manifolds by the spin multiplicity defined by
\begin{align}
\text{multiplicity} = 2S + 1
\end{align}
where the eigenstates of the $S^2$ operator have, by definition, eigenvalues of $S(S+1)$.  In our example, we have used a singlet state with $S=0$ to specify we are looking for the lowest singlet energy state.  This was
specified simply by
\begin{lstlisting}[language=Python]
multiplicity = 1
\end{lstlisting}
and we note that a multiplicity of $3$ (a triplet state) might have also been of interest for this particular system.

Given a particular set of nuclei, it is often the case that a system with the same number of electrons as protons
(or electronically neutral system) is the most stable.  However, sometimes one wishes to study systems with a
non-neutral charge such as the $+1$ cation or $-1$ anion of the system.  In that case, one may specify the charge, which
is defined to be the number of electrons in the neutral system minus the number of electrons in the system of interest.
In our example, we specified the neutral hydrogen atom, with
\begin{lstlisting}[language=Python]
charge = 0
\end{lstlisting}

Finally, to specify the computational problem to be solved, rather than simply the molecule itself, it is necessary
to define the basis set.  This is analogous to, in some loose sense, how one might define a grid to solve a
a simple partial differential equation such as the heat-equation.  In chemistry, much thought has gone into optimizing specialized basis sets
to pack the essential physics of the problems into functions that balance cost and accuracy.  A list
of some of the most common basis sets expressed as sums of Gaussians can be found in the EMSL basis set exchange~\cite{Schuchardt:2007}.
In this case, we specify the so-called minimal basis set ``sto-3g'', which stands for 3 Gaussians (3G) used to approximate
Slater-type orbitals (STO).  This is done through the line 
\begin{lstlisting}[language=Python,upquote=True]
basis = 'sto-3g'
\end{lstlisting}
In future implementations, the code may additionally support parametric or user defined basis sets with similar syntax.
At present, the code supports basis sets that are implemented within common molecular electronic structure packages
that will be described in more detail in the next section.  With the above specifications, the chemical problem is now well defined; however, several steps remain in mapping to a qubit representation.

\subsection{Integral generation}

After the specification of the molecule as in the previous section, it becomes necessary to do some of the numerical
work in generating the problem to be solved.   In OpenFermion we accomplish this through plugin libraries that interface with
existing electronic structure codes. At present there are supported interfaces to Psi4~\cite{Parrish:2017} and PySCF~\cite{Sun:2017}, with plans
to expand to other common packages in the future. One needs to install these plugin libraries separately from the core OpenFermion library (instructions are provided at \href{http://www.openfermion.org}{\texttt{www.openfermion.org}} that includes a Docker alternative installing all these packages). We made the decision to support these packages as plugins rather than directly integrating them for several reasons. First, some of these packages have different open source licenses than OpenFermion. Second, these packages may require more intricate installations and do not run on all operating systems. Finally, in the future one may wish to use OpenFermion in conjunction with other electronic structure packages which might support methods not implemented in Psi4 and PySCF. The plugin model ensures the modularity necessary to maintain such compatibilities as the code evolves.

Once one has chosen a particular basis and enforced the physical
anti-symmetry of electrons, the electronic structure problem may be written exactly in the form of a second quantized electronic Hamiltonian
as
\begin{align}
H(R) = \sum_{ij} h_{ij}(R) a_i^\dagger a_j + \frac{1}{2} \sum_{ijkl} h_{ijkl}(R) a_i^\dagger a_j^\dagger a_k a_l.
\end{align}
The coefficients $h_{ij}$ and $h_{ijkl}$ are defined by the basis set that was chosen for the problem (sto-3g in our example);
however the computation of these coefficients in general can be quite involved.  In particular, in many cases it
makes sense to perform a Hartree-Fock calculation first and transform the above integrals into the molecular
orbital basis that results.  This has the advantage of making the mean-field state easy to represent on both
a classical and quantum computer, but introduces some challenges with regards to cost of the integral transformation
and convergence of the Hartree-Fock calculation for challenging systems.  For example, it is necessary to
specify an initial guess for the orbitals of Hartree-Fock, the method used to solve the equations, and a number
of other numerical parameters that affect convergence.

In OpenFermion we address this problem by choosing a reasonable set of default parameters and interfacing with
well developed electronic structure packages through a set of plugin libraries to supply the information desired.
For example, using the OpenFermion-Psi4 plugin, one can obtain the two-electron integrals $h_{ijkl}$ for this molecule in the MO basis in the Psi4 electronic structure code by simply executing
\begin{lstlisting}[language=Python]
from openfermionpsi4 import run_psi4

h2_molecule = run_psi4(h2_molecule,
                       run_mp2=True,
                       run_cisd=True,
                       run_ccsd=True,
                       run_fci=True)
two_electron_integrals = h2_molecule.two_body_integrals
\end{lstlisting}
where the h2\_molecule is that defined in the previous section and when run in this way, one may easily access the MP2, CISD, CCSD, and FCI energies.  This will return the computed two-electron integrals in the Hartree-Fock molecular orbital basis.  Moreover,
one has direct access to other common properties such as the molecular orbital coefficients through 
simple commands such as
\begin{lstlisting}[language=Python]
orbitals = h2_molecule.canonical_orbitals
\end{lstlisting}
One may also read the $1$- 
and $2$-electron reduced density matrices from CISD and FCI and the converged coupled cluster amplitudes from CCSD.
These values are all conveniently stored to disk using an HDF5 interface that only loads the
properties of interest from disk for convenient analysis of data after the fact. 

The plugins are designed to ideally function uniformly without exposing the details of the underlying package to the user.  For example, the same computation may be accomplished using PySCF through the OpenFermion-PySCF plugin by executing the commands
\begin{lstlisting}[language=Python]
from openfermionpyscf import run_pyscf
                                       
h2_molecule = run_pyscf(h2_molecule,
                        run_mp2=True,
                        run_cisd=True,
                        run_ccsd=True,
                        run_fci=True)
                        
h2_filename = h2_molecule.filename                        
h2_molecule.save()
\end{lstlisting}
This allows the user to prepare a data structure representing the molecule that is agnostic to the package with which it was generated.  In the future, additional plugins will support a wider range of electronic structure packages to meet the growing demands of users.  In the last line of this example, we introduce the save feature of the MolecularData class.  We note that this is called by default by the plugins generating the data, but introduce it here to emphasize the fact that this data is conveniently stored for future use.  This allows one to retrieve any data about this molecule in the future without an additional calculation through
\begin{lstlisting}[language=Python]
from openfermion.hamiltonians import MolecularData
                                       
h2_molecule = MolecularData(filename=h2_filename)
\end{lstlisting}
where h2\_filename is the filename one chose to store the H$_2$ data under.  By using decorated Python attributes and HDF5 storage, the loading of data in this way is done on-demand for any array-like object such as integrals.  That is, loading the molecule in this way has a minimal memory footprint, and large quantities such as density matrices or integrals are only loaded when the attribute is accessed, for example
\begin{lstlisting}[language=Python]
one_body_integrals = h2_molecule.one_body_integrals
\end{lstlisting}
and is performed seamlessly in the background such that no syntax is required beyond accessing the attribute. 

\subsection{Mapping to qubits}
After the problem has been recast in the second quantized representation, it remains to map the problems to qubits.  Electrons are anti-symmetric indistinguishable particles, while qubits are distinguishable particles, so some care must be taken in mapping between the two.  There are now many maps that respect the correct particle statistics, and several
of the most common are currently implemented within OpenFermion.  In particular, the Jordan-Wigner (JW)~\cite{Jordan1928}, 
Bravyi-Kitaev (BK)~\cite{Bravyi2002,Seeley2012}, and Bravyi-Kitaev super fast (BKSF)~\cite{Setia2017} transformations are currently supported in OpenFermion.  Each of these has different properties
with regard to the Hamiltonians that are produced, which may offer benefits to different types of algorithms or experiments.  OpenFermion attempts to remain agnostic to the particular transformation preferred by the user.

To give a concrete example, the above Hamiltonian may be mapped to a qubit Hamiltonian through the use of the
Jordan-Wigner transformation by
\begin{lstlisting}[language=Python]
from openfermion.transforms import get_fermion_operator, jordan_wigner

h2_qubit_hamiltonian = jordan_wigner(get_fermion_operator(h2_molecule.get_molecular_hamiltonian()))
\end{lstlisting}
which returns a qubit operator representing the Hamiltonian after the Jordan-Wigner transformation.  This data structure
will be explored in more detail later in this paper, but it provides the complete specification of the Hamiltonian
acting on qubits in a convenient format.

Note that, in addition to mapping fermionic Hamiltonians to qubit operators, OpenFermion also supports mapping bosonic Hamiltonians --- those describing systems of indistinguishable symmetric particles --- to propagation modes of light, or \textit{qumodes}, through the use of plugins to quantum optics-based simulators. In future iterations of OpenFermion, algorithms mapping bosonic systems to qubits may additionally be implemented.

\subsection{Numerical testing}
While the core functionality of OpenFermion is designed to provide a map between the space of electronic structure
problems and qubit-based quantum computers, some functionality is provided for numerical simulation.  This can be
helpful for debugging or prototyping new algorithms.  For example, one could check the spectrum of the
above qubit Hamiltonian after the Jordan-Wigner transformation by performing
\begin{lstlisting}[language=Python]
from openfermion.transforms import get_sparse_operator
from scipy.linalg import eigh

h2_matrix = get_sparse_operator(h2_qubit_hamiltonian).todense()
eigenvalues, eigenvectors = eigh(h2_matrix)
\end{lstlisting}
which yields the exact eigenvalues and eigenvectors of the H$_2$ Hamiltonian in the computational basis.

\subsection{Compiling circuits for quantum algorithms}

The core of OpenFermion consists of tools for obtaining and manipulating representations of quantum operators. These tools and representations are useful for compiling quantum circuits, but the best way to perform this compilation depends on the particular algorithm and hardware used. Many groups have their own software for performing this compilation. OpenFermion is intended to be agnostic with respect to the particular hardware and compilation platform; we delegate the final compilation step to platform-specific plugins. At the time of writing, plugins are supported for the Cirq, Forest, and Strawberry Fields~\cite{Killoran2018} frameworks. These plugins are called OpenFermion-Cirq, Forest-OpenFermion, and SFOpenBoson, respectively. Below, we provide examples of using these plugins to compile quantum circuits.

\subsubsection{OpenFermion-Cirq Example}

In our first example, we assume the user has a hydrogen molecule object stored in the variable \emph{h2\_molecule} and use Cirq and OpenFermion-Cirq to construct a circuit that simulates time evolution by the molecular Hamiltonian via a product formula implemented with the ``low rank'' strategy described in \cite{Motta2018}. We'll compile just one Trotter step and perform a circuit optimization which merges single-qubit rotations before displaying the circuit.

\begin{lstlisting}[language=Python]
import cirq
import openfermioncirq as ofc

hamiltonian = h2_molecule.get_molecular_hamiltonian()
qubits = cirq.LineQubit.range(4)
circuit = cirq.Circuit.from_ops(
    ofc.simulate_trotter(
        qubits,
        hamiltonian,
        time=1.0,
        n_steps=1,
        order=0,
        algorithm=ofc.trotter.LOW_RANK,
        omit_final_swaps=True
    )
)
cirq.merge_single_qubit_gates_into_phased_x_z(circuit)

print(circuit.to_text_diagram(use_unicode_characters=False))
\end{lstlisting}

Executing this code produces an ASCII circuit diagram which we truncate and display below:

\begin{Verbatim}[fontsize=\small]
0: ---Z^0.535-------------------YXXY--------------------@----------swap--------------------------
                                |                       |          |
1: ---Z^0.535-----------YXXY----#2^-1-----------YXXY----@^-0.218---swap--------------@-----------
                        |                       |                                    |             ...
2: ---Z^0.291-----------#2^-1-----------YXXY----#2^-1--------------@----------swap---@^(-3/14)---
                                        |                          |          |
3: ---Z^0.291---------------------------#2^-1----------------------@^-0.211---swap---------------
\end{Verbatim}

In our next example, we construct a circuit which prepares the ground state of the tunneling term
of a Fermi-Hubbard model Hamiltonian. The Fermi-Hubbard model has the Hamiltonian
$$
-t \sum_{\langle i, j \rangle, \sigma} (a_{i, \sigma}^\dagger a_{j, \sigma} + a_{j, \sigma}^\dagger a_{i, \sigma})
    + U \sum_{i} a_{i, \uparrow}^\dagger a_{i, \uparrow} a_{i, \downarrow}^\dagger a_{i, \downarrow} \\
$$
which is composed of a tunneling term (on the left) and an interaction term (on the right).
The tunneling term is a quadratic Hamiltonian and its ground state is a Slater determinant.
More generally, eigenstates of quadratic Hamiltonians are known as fermionic Gaussian states.

\begin{lstlisting}[language=Python]
import cirq
import openfermion
import openfermioncirq as ofc

hubbard_model = openfermion.fermi_hubbard(2, 2, 1.0, 4.0)
quad_ham = openfermion.get_quadratic_hamiltonian(hubbard_model, ignore_incompatible_terms=True)

qubits = cirq.LineQubit.range(8)
circuit = cirq.Circuit.from_ops(
    ofc.prepare_gaussian_state(qubits, quad_ham)
)

print(circuit.to_text_diagram(transpose=True, use_unicode_characters=False))
\end{lstlisting}

Executing this code produces the following ASCII circuit diagram:

\begin{Verbatim}[fontsize=\small]
0    1        2         3         4          5        6         7
|    |        |         |         |          |        |         |
X    X        X         X         |          |        |         |
|    |        |         |         |          |        |         |
|    |        |         YXXY------#2^0.859   |        |         |
|    |        |         |         |          |        |         |
|    |        YXXY------#2^-0.701 Z^0        |        |         |
|    |        |         |         |          |        |         |
|    YXXY-----#2^-0.89  Z^0       YXXY-------#2^0.805 |         |
|    |        |         |         |          |        |         |
YXXY-#2^0.844 Z^0       YXXY------#2^-0.859  Z^0      |         |
|    |        |         |         |          |        |         |
|    Z^0      YXXY------#2^0.607  Z^0        YXXY-----#2^0.844  |
|    |        |         |         |          |        |         |
|    YXXY-----#2^-0.569 Z^0       YXXY-------#2^0.565 Z^0       |
|    |        |         |         |          |        |         |
|    |        Z^0       YXXY------#2^(13/15) Z^0      YXXY------#2^0.675
|    |        |         |         |          |        |         |
|    |        YXXY------#2^-0.853 Z^0        YXXY-----#2^-0.565 Z^0
|    |        |         |         |          |        |         |
|    |        |         Z^0       YXXY-------#2^0.893 Z^0       |
|    |        |         |         |          |        |         |
|    |        |         YXXY------#2^0.381   Z^0      |         |
|    |        |         |         |          |        |         |
|    |        |         |         Z^0        |        |         |
|    |        |         |         |          |        |         |
\end{Verbatim}

\subsubsection{Forest-OpenFermion Example}
In this section we describe the interface between OpenFermion and Rigetti's quantum simulation environment called Forest. The interface provides a method of transforming data generated in OpenFermion to a similar representation in pyQuil.  For this example we use OpenFermion to build a four-site single-band periodic boundary Hubbard model and apply first-order Trotter time-evolution to a starting state of two localized electrons of opposite spin.

The Forest-OpenFermion plugin provides the routines to inter-convert between the OpenFermion QubitOperator data structure and the synonymous data structure in pyQuil called a PauliSum.  
\begin{lstlisting}[language=Python]
from openfermion.ops import QubitOperator
from forestopenfermion import pyquilpauli_to_qubitop, qubitop_to_pyquilpauli
\end{lstlisting}
The FermionOperator in OpenFermion can be used to translate the mathematical expression of the Hamiltonian directly to executable code.  While we show how this model can be built in one line using the OpenFermion hamiltonians module below, here we take the opportunity to demonstrate the ease of creating such models for study that could be easily modified as desired.  Given the Hamiltonian of the Hubbard system
\begin{align}
H = -1\sum_{\langle i, j \rangle, \sigma}\left( a_{i, \sigma}^{\dagger}a_{j, \sigma} + a_{j, \sigma}^{\dagger}a_{i} \right) + 4 \sum_{i=0}^{3} a_{i, \alpha}^{\dagger}a_{i, \alpha}a_{i, \beta}^{\dagger}a_{i, \beta}
\end{align}
where $\langle \cdot, \cdot \rangle$ indicates nearest-neighbor spatial lattice positions and $\sigma$ takes on values of $\alpha$ and $\beta$ signifying spin-up or spin-down, respectively, the code to build this Hamiltonian is as follows:
\begin{lstlisting}[language=Python]
from openfermion.transforms import jordan_wigner
from openfermion.ops import FermionOperator, hermitian_conjugated

hubbard_hamiltonian = FermionOperator()
spatial_orbitals = 4
for i in range(spatial_orbitals):
    electron_hop_alpha = FermionOperator(((2 * i, 1), (2 * ((i + 1) % spatial_orbitals), 0)))
    electron_hop_beta = FermionOperator(((2 * i + 1, 1), ((2 * ((i + 1) % spatial_orbitals) + 1), 0)))
    hubbard_hamiltonian += -1 * (electron_hop_alpha + hermitian_conjugated(electron_hop_alpha))
    hubbard_hamiltonian += -1 * (electron_hop_beta + hermitian_conjugated(electron_hop_beta))
    hubbard_hamiltonian += FermionOperator(((2 * i, 1), (2 * i, 0),
                                            (2 * i + 1, 1), (2 * i + 1, 0)), 4.0)
\end{lstlisting}
In the above code we have implicitly used even indexes $[0, 2, 4, 6]$ as $\alpha$ spin-orbitals and odd indexes $[1, 3, 5, 7]$ as $\beta$ spin-orbitals.  The same model can be built using the OpenFermion Hubbard model builder routine in the hamiltonians module  with a single function call.
\begin{lstlisting}[language=Python]
from openfermion.hamiltonians import fermi_hubbard

x_dim = 4
y_dim = 1
periodic = True
chemical_potential = 0
tunneling = 1.0
coulomb = 4.0
of_hubbard_hamiltonian = fermi_hubbard(x_dim, y_dim, tunneling, coulomb,
                                       chemical_potential=None,
                                       spinless=False)
\end{lstlisting}
Using the Jordan-Wigner transform functionality of OpenFermion, the Hubbard Hamiltonian can be transformed to a sum of QubitOperators which are then transformed to pyQuil PauliSum objects using routines in the Forest-OpenFermion plugin imported earlier.
\begin{lstlisting}[language=Python]
hubbard_term_generator = jordan_wigner(hubbard_hamiltonian)
pyquil_hubbard_generator = qubitop_to_pyquilpauli(hubbard_term_generator)
\end{lstlisting}
With the data successfully transformed to a pyQuil representation, the pyQuil exponentiate routine is used to generate a circuit corresponding to first-order Trotter evolution for $t=0.1$.
\begin{lstlisting}[language=Python]
from pyquil.quil import Program
from pyquil.gates import X
from pyquil.paulis import exponentiate
localized_electrons_program = Program()
localized_electrons_program.inst([X(0), X(1)])
pyquil_program = Program()
for term in pyquil_hubbard_generator.terms:
    pyquil_program += exponentiate(0.1 * term)
print(localized_electrons_program + pyquil_program)
\end{lstlisting}
\begin{lstlisting}[language=Python]
Sample Output:
X 0
X 1
X 0
PHASE(-0.4) 0
X 0
PHASE(-0.4) 0
H 0
H 2
CNOT 0 1
CNOT 1 2
RZ(-0.1) 2
CNOT 1 2
CNOT 0 1
...
\end{lstlisting}
The output is the first few lines of the Quil \cite{Smith:2016} program that sets up the two-localized electrons on the first spatial site and then applies the time-propagation circuit.  This Quil program can be sent to the Forest cloud API for simulation or for execution on hardware.

\subsubsection{SFOpenBoson Example}
Quantum optics and continuous-variable (CV) quantum computation provide a well-suited environment for the quantum simulation of bosons. In this example, we describe the interface between OpenFermion and Xanadu's quantum optics-based simulation environment Strawberry Fields, and simulate a Bose-Hubbard model using the CV gate set and propagation modes of light.

Consider the following Bose-Hubbard Hamiltonian, describing bosons on a lattice with on-site interaction strength $U=1.5$ and chemical potential $\mu=0.5$:

\begin{align}
	H = - \sum_{\langle i, j \rangle} b_i^\dagger b_{j + 1}
	+ \frac{1.5}{2} \sum_{k=1}^{N-1} b_k^\dagger b_k (b_k^\dagger b_k - 1)
	- 0.5 \sum_{k=1}^N b_k^\dagger b_k.
\end{align}
Here, $b_i^\dagger$ and $b_i$ are the bosonic creation and annhilation operators respectively acting on qumode $i$. Using OpenFermion, it is a one line procedure to generate this Hamiltonian for a $2\times 2$ lattice:
\begin{lstlisting}[language=Python]
from openfermion.hamiltonians import bose_hubbard
H = bose_hubbard(2, 2, tunneling=1, interaction=1.5, chemical_potential=0.5)
\end{lstlisting}
We can now use the SFOpenBoson plugin to simulate this Hamiltonian in Strawberry Fields. The SFOpenBoson plugin provides operations to convert the OpenFermion BosonOperator/QuadOperator data structure to the standard operations and gates used in continuous-variable quantum computation. Using the Hamiltonian defined above, we can do this as follows:
\begin{lstlisting}[language=Python]
import strawberryfields as sf
from strawberryfields.ops import Fock
from sfopenboson.ops import BoseHubbardPropagation

t = 0.1
eng, q = sf.Engine(4)
with eng:
    Fock(2) | q[0]
    BoseHubbardPropagation(H, t, k=20) | q
\end{lstlisting}
In the above code, we create a two qumode quantum circuit with the system initialized in Fock state $\left|2,0,0,0\right\rangle$, and then perform the time evolution for $t=0.1$ using the Lie product formula truncated to $20$ terms. Finally, we run the circuit simulation, and print the applied quantum gates:
\begin{lstlisting}[language=Python]
state = eng.run('fock', cutoff_dim=3)
eng.print_applied()
\end{lstlisting}
\begin{lstlisting}[language=Python]
BSgate(-0.005, 1.571) | (q[0], q[1]) # Beamsplitter
BSgate(-0.005, 1.571) | (q[0], q[2]) # Beamsplitter
BSgate(-0.005, 1.571) | (q[1], q[3]) # Beamsplitter
BSgate(-0.005, 1.571) | (q[2], q[3]) # Beamsplitter
Kgate(-0.00375) | (q[0]) # Kerr interaction
...
\end{lstlisting}
The output above lists the first few gates applied, and the corresponding qumodes they act on, for the Bose-Hubbard simulation.

\section{Data structures}
In this section we describe some of the data structures defined by OpenFermion. We begin
with the most general data structures: FermionOperator, QubitOperator, and BosonOperator.
These represent arbitrary linear combinations of operators from the chosen basis; for
instance, for FermionOperator, the basis is that of fermionic creation and annihilation operators.
Later on we describe data structures such as InteractionOperator which represent
operators with additional structure that warrants storage in a different format.

\subsection{FermionOperator data structure}
\label{sec:fermion_operator}

In the above examples, we saw that an intermediate representation for the molecular problems were objects known as FermionOperators.
Fermionic systems are often treated in second quantization, where anti-symmetry requirements are stored in the operators rather than in
explicit wavefunction anti-symmetrization and arbitrary operators can be expressed using fermionic creation and annihilation operators
$a^\dagger_p$ and $a_q$. Supposing that $p=1$, $q=0$, such operators could be represented within OpenFermion simply as
\begin{lstlisting}[language=Python,upquote=true]
from openfermion.ops import FermionOperator
a_p_dagger = FermionOperator('1^')
a_q = FermionOperator('0')
\end{lstlisting}
These operators enforce fermionic statistics in the system by satisfying the fermionic anti-commutation relations,
\begin{equation}
\label{eq:commutation}
\left\{a^\dagger_p, a^\dagger_q\right\} = \left\{a_p, a_q\right\} = 0 \quad \quad \left\{a^\dagger_p, a_q\right\} = \delta_{pq}.
\end{equation}
where $\{A,B\} \equiv AB + BA$. The raising operators $a_p^\dagger$ act on the fermionic vacuum state, 
$\ket{vac}$, to create fermions in spin-orbitals, which are single-particle spatial density functions.  The connection to first
quantization and explicit anti-symmetrization in Slater determinants can be seen if electron 
$j$ is represented in a space of spin-orbitals $\{\phi_p(r_j)\}$. Then $a^\dagger_p$ and $a_p$  populate fermions in Slater
determinants through the equivalence,
\begin{align}
&  \bra{r_0,\ldots,r_{\eta-1}} a^\dagger_{p_0} \cdots a^\dagger_{p_{\eta-1}} \ket{vac} = \sqrt{\frac{1}{\eta!}}
\begin{vmatrix}
\phi_{p_0}\left(r_0\right) & \phi_{p_1}\left( r_0\right) & \cdots & \phi_{p_{\eta-1}} \left( r_0\right) \\
\phi_{p_0}\left(r_1\right) & \phi_{p_1}\left( r_1\right) & \cdots & \phi_{p_{\eta-1}} \left( r_1\right) \\
\vdots & \vdots & \ddots & \vdots\\
\phi_{p_0}\left(r_{\eta-1}\right) & \phi_{p_1}\left(r_{\eta-1}\right) & \cdots & \phi_{p_{\eta-1}} \left(r_{\eta-1}\right) \end{vmatrix}
\end{align}
which instantiates a system of $\eta$ fermions.

Arbitrary fermionic operators on the space of $N$ spin-orbitals can be represented by weighted sums of products of these raising and lowering operators. The following is an example of one such ``fermion operator'',
\begin{equation}
W = \left(1 + 2 i\right) a^\dagger_4 a_3 a_9 a^\dagger_3 - 4 \, a_2 = -\left(1 + 2 i\right) a^\dagger_4 a_3^\dagger a_9 a_3 -\left(1 + 2 i\right) a^\dagger_4 a_9 - 4 \, a_2.
\end{equation}
In the second equality above we have used the anti-commutation relations of \eq{commutation} to reorder the ladder operators in $W$ into a unique ``normal-ordered'' form, defined so that raising operators always come first and operators are ordered in descending order of the fermionic mode on which they act. These rules are all handled transparently within OpenFermion so that essential physics are not violated.  For example, the $W$ operator could be defined within OpenFermion as
\begin{lstlisting}[language=Python,upquote=true]
from openfermion.ops import FermionOperator
W = (1 + 2j) * FermionOperator('4^ 3 9 3^') - 4 * FermionOperator('2')
\end{lstlisting}
and the ``normal-ordering'' can be simply performed by
\begin{lstlisting}[language=Python]
from openfermion.utils import normal_ordered
W_normal_ordered = normal_ordered(W)
\end{lstlisting}
So long as ladder operators are manipulated in a fashion that is consistent with \eq{commutation}, addition, multiplication, and integer exponentiation are well defined for fermion operators. For instance, $W^4$ and $W / 82 - 3 W^2$ are also examples of fermion operators,
and are readily available in OpenFermion through standard arithmetic manipulations of the operators.  For example
\begin{lstlisting}[language=Python]
W_4 = W ** 4
print(normal_ordered(W_4))
\end{lstlisting}
where in the second line, we find if we further use the normal ordered function on this seemingly complicated object, that it in fact evaluates to zero.

Internally, the FermionOperator data structure uses a hash table (currently implemented using a Python dictionary). The keys of the dictionary encode a sequence of raising and lowering operators and the value of that entry stores the coefficient. The current implementation of this class encodes the sequence of ladder operators using a tuple of 2-tuples where the 2-tuples represent ladder operators. The first element of each 2-tuple is an int specifying which fermionic mode the ladder operator acts on and the second element of each 2-tuple is a Boolean specifying whether the operator is raising (True) or lowering (False); thus, the encoding of the ladders operators can be expressed as
\begin{equation}
(p \in \mathbb{N}_0, \gamma \in \mathbb{Z}_2 ) \mapsto \begin{cases} a^\dagger_p & \gamma = 1\\ a_p & \gamma = 0 \end{cases}.
\end{equation}
The sequence of ladder operators is thus specified by a sequence of the 2-tuples just defined. Some examples of the ladder operator sequence encodings are shown below,
\begin{equation}
() \mapsto \openone \quad \quad ((2, 0),) \mapsto a_2  \quad \quad  ((4, 1), (9, 0)) \mapsto a^\dagger_4 a_9 \quad \quad ((4, 1), (3, 0), (9, 0), (3, 1)) \mapsto a^\dagger_4 a_3 a_9 a_3^\dagger.
\end{equation}
which can also be used to initialize a FermionOperator as a user as
\begin{lstlisting}[language=Python]
O_1 = FermionOperator()
O_2 = FermionOperator( ((2, 0), ) )
O_3 = FermionOperator( ((4, 1), (9, 0)) )
O_4 = FermionOperator( ((4, 1), (3, 0), (9, 0), (3, 1)) )
\end{lstlisting}

While this is the internal representation of sequences of ladder operators in the FermionOperator data structure, OpenFermion also supports a string representation of ladder operators that is more human-readable. One can initialize FermionOperators using the string representation and when one calls \verb|print()| on a FermionOperator, the operator is printed out using the string representation. The carat symbol ``\,\^ \,'' represents raising, and its absence implies the lowering operator. Below are some self-explanatory examples of our string representation,
\begin{equation}
\textrm{{\ttfamily\char'15}}\textrm{{\ttfamily\char'15}} \mapsto \openone
\quad \quad
\textrm{{\ttfamily\char'15}}2\textrm{{\ttfamily\char'15}} \mapsto a_2
\quad \quad 
\textrm{{\ttfamily\char'15}}4\textrm{\textasciicircum} \,\, 9\textrm{{\ttfamily\char'15}} \mapsto a^\dagger_4 a_9
\quad \quad 
\textrm{{\ttfamily\char'15}}4\textrm{\textasciicircum} \,\, 3 \,\, 9 \,\, 3\textrm{\textasciicircum}\textrm{{\ttfamily\char'15}} \mapsto a^\dagger_4 a_3 a_9 a_3^\dagger.
\end{equation}
that translate to code as
\begin{lstlisting}[language=Python,upquote=true]
O_1 = FermionOperator('')
O_2 = FermionOperator('2')
O_3 = FermionOperator('4^ 9')
O_4 = FermionOperator('4^ 3 9 3^')
\end{lstlisting}
A hash table data structure was chosen to facilitate efficient combination of large FermionOperators through arithmetic operations. 
This is preferred over an unstructured array of terms due to its native implementation in Python as well as the fact that duplicate terms
are nearly automatically combined at a cost that is constant time for modestly sized examples.  Similar scaling could be achieved through
other data structures, but at increased complexity of implementation in the Python ecosystem.  As motivation for this choice, we include in the
example section two important use cases of the FermionOperator: the computation of Trotter error operators and the symbolic Fourier transformation.

\subsection{QubitOperator data structure}
Continuing from the example above, once the intermediate FermionOperators have been produced, they must be mapped to the language of
quantum computers, or qubits.  This is handled within OpenFermion through the QubitOperator data structure.  Fundamentally this operator
structure is based off the Pauli spin operators and the identity, defined by
\begin{align}
I = \left( \begin{array}{cc}
     1 & 0  \\
     0 & 1
\end{array} \right) \quad \quad
X = \left( \begin{array}{cc}
     0 & 1  \\
     1 & 0
\end{array} \right) \quad \quad
Y = \left( \begin{array}{cc}
     0 & -i  \\
     i & 0
\end{array} \right) \quad \quad
Z = \left( \begin{array}{cc}
     1 & 0  \\
     0 & -1
\end{array} \right).
\end{align}
Tensor products of Pauli operators form a basis for the space of Hermitian operators; thus, it is possible to express any Hermitian
operator of interest using just these few operators and their products.  If one indexes the qubit a particular operator 
acts on, such as $X_i$, that defines action by $X$ on the $i^\text{th}$ qubit (and implicitly action by $I$ on all others).  In OpenFermion
one may wish to express an operator such as
\begin{align}
O = Z_1 Z_2 + X_1 + X_2
\end{align}
that could be initialized as
\begin{lstlisting}[language=Python,upquote=true]
from openfermion.ops import QubitOperator
O = QubitOperator('Z1 Z2') + QubitOperator('X1') + QubitOperator('X2')
\end{lstlisting}
Similar to FermionOperator, QubitOperator is implemented internally using a hash table data structure through the native Python
dictionary.  This choice allows a good level of base efficiency for most arithmetic operators while harnessing the features
of the Python dictionary to ease implementation details.  The keys used in this implementation are similarly tuples of tuples that define
the type of operator and the qubit it acts on,
\begin{align}
(O \in \{X, Y, Z\}, i \in \mathbb{N}_0 ) \mapsto O_i.
\end{align}
This internal representation may be used to initialize QubitOperators such as the operator $X_1 X_2$ as
\begin{lstlisting}[language=Python,upquote=true]
O = QubitOperator( ((1, 'X'), (2, 'X')) )
\end{lstlisting}
or alternatively as seen above, a convenient string initializer is also available, which for the same operator could be used as
\begin{lstlisting}[language=Python,upquote=true]
O = QubitOperator('X1 X2')
\end{lstlisting}

\subsection{BosonOperator and QuadOperator data structures}
In addition to the FermionOperator, OpenFermion also provides preliminary support for bosonic systems through the BosonOperator data structure. Like with fermionic systems, bosonic systems are commonly treated in second quantization, using the symmetric bosonic creation and annihilation operators $b_q^\dagger$ and $b_q$. These enforce bosonic statistics by satisfying the canonical bosonic commutation relations,
\begin{align}
	[a_p^\dagger, a_q^\dagger] = [a_p, a_q] = 0, ~~~~ [a_p^\dagger, a_q] = \delta_{pq},
\end{align}
and, analogously to the fermionic case, act on the vacuum state to create/annihilate bosons in the Fock space:
\begin{align}
	{a_p^\dagger}^n\ket{vac} = \ket{\underbrace{0, 0, \dots, 0}_{{p-1}}, n, 0, \dots, 0}.
\end{align}
Due to these similarities, the BosonOperator is inherited from the same SymbolicOperator class as the FermionOperator, and thus retains many of the same properties, including the internal representation through the use of a hash table, and forms of input (terms can be specified as a 2-tuple or in the more human-readable string representation). For example, constructing the operator $b_3^\dagger b_5 b_1^\dagger b_4$ can be done by the user as
\begin{lstlisting}[language=python]
from openfermion.ops import BosonOperator
O = BosonOperator('3^ 5 1^ 4')
\end{lstlisting}
Note that, unlike fermionic ladder operators, bosonic ladder operators acting on different subsystems commute past each other. Therefore, the terms of this operator will be stored as ``1\^{} 3\^{} 4 5'', with indices ordered in ascending order from left to right. To recover the unique ``normal-ordered'' form, with raising operators always to the left of lowering operators, the user can use the normal ordered function provided to automatically rearrange the operator terms as per the commutation relations:
\begin{lstlisting}[language=python]
from openfermion.utils import normal_ordered
normal_ordered(BosonOperator('0 0^'))
\end{lstlisting}
which returns ``1.0 [~] + 1.0 [0\^{} 0]'', corresponding to $b_0 b_0^\dagger = I+b_0^\dagger b_0$.

In addition to bosonic raising and lowering operators, it is common when studying bosonic systems to also consider dynamics in the phase space $(q,p)$. The Hermitian position and momentum operators $q_i$ and $p_i$ act upon this phase space, and are defined in terms of the bosonic ladder operators,
\begin{align}
	q_i = \sqrt{\frac{\hbar}{2}}\left(b_i+b_i^\dagger\right), ~~~p_i =-i \sqrt{\frac{\hbar}{2}}\left(b_i-b_i^\dagger\right).
\end{align}
These operators are Hermitian, and are referred to as the quadrature operators. They satisfy the commutation relation $[q_i,p_j]=i\hbar\delta_{ij}$, where $\hbar$ depends on convention, often taking the values $\hbar=0.5,1,$ or $2$.

In OpenFermion, the quadrature operators are represented by the QuadOperator class, and stored as a dictionary of tuples (as keys) and coefficients (as values), similar to the QubitOperator. For example, the multi-mode quadrature operator $q_0 p_1 q_3$ is represented internally as \texttt{((0, `q'), (1, `p'), (3, `q'))}. Alternatively, QuadOperators also support string input --- to initialize the latter operator using string representation, the user would enter the following:
\begin{lstlisting}[language=python]
from openfermion.ops import QuadOperator
O = QuadOperator('q0 p1 q3')
\end{lstlisting}

As with the BosonOperator, quadrature operators acting on different subsystems commute, so the internal representation orders the terms with lowest index to the left. We also define a ``normal-ordering'' for quadrature operators, to allow a unique representation. The normal-ordered function introduced earlier rearranges the terms via the commutation relation such that position operators $q_i$ always appear to the left of momentum operators $p_i$:
\begin{lstlisting}[language=python]
normal_ordered(QuadOperator('p0 q0'), hbar=2.)
\end{lstlisting}
Note that we now also pass the value $\hbar$ that appears in the commutation relation; if not provided, the default is $\hbar=1$.

Finally, OpenFermion provides functions for converting between the BosonOperator and QuadOperator:
\begin{lstlisting}[language=python]
from openfermion.transforms import get_quad_operator, get_boson_operator
O = QuadOperator('q0 p1 q3')
O2 = get_boson_operator(O, hbar=2.)
\end{lstlisting}
As before, we pass the value of $\hbar$ required depending on the convention chosen. 

In addition to the functions and methods shown here, the BoseOperator and QuadOperator support additional operations (for example, sparse operator representations, symmetric ordering, and Weyl quantization of observables in the phase space). They also form the basis for models provided by OpenFermion, such as the Bose-Hubbard Hamiltonian.

\subsection{The MolecularData data structure}

While the FermionOperator, BosonOperator, and QubitOperator classes form the backbone of many of the internal computations for OpenFermion, the data
that defines a particular physical problem is more conveniently stored within a separate well-defined object,  namely, the MolecularData
object.  This defines the schema by which the intermediate quantities calculated for the electronic structure of a molecule are stored,
such as the two-electron integrals, basis transformation from the original basis, energies from correlated methods, and meta data
related to the computation.  For an exhaustive list of the current information stored within this class, it is recommended to see
the documentation, as quantities are added as needed.

Importantly, this information is stored to disk in an HDF5 format using the \verb|h5py| package.  This allows for seamless data
access to the files for only the quantities of interest, without the need for loading the whole file or complicated interface
structures.  Internally this is performed through the use of getters and setters using Python decorators, but this is transparent to
the user, needing only to get or set in the normal way,
e.g.
\begin{lstlisting}[language=Python]
two_body_integrals = h2_molecule.two_body_integrals
\end{lstlisting}
where the data is read from disk on the access to two\_body\_integrals rather than when the object is instantiated in the first line.
This controls the memory impact of larger objects such as the two-electron integrals.  Compression functionality is also enabled through
\verb|gzip| to minimize the disk space requirements to store larger molecules.

Functionality is also built into MolecularData class to perform simple activate space approximations to the problem.  An active space approximation isolates a subset of the total orbitals and treats the problem within that orbital set.  This is done by modifying the one- and two-electron integrals as well as the count of active electrons and orbitals within the molecule.  We assume a single reference for the active space definition and the occupied orbitals are integrated out according to the integrals while inactive virtual orbitals are removed.  In OpenFermion this is as simple as taking a calculated molecule data structure, forming a list of the occupied spatial orbital indices (those which will be frozen in place) and active spatial orbital indices (those which may be freely occupied or unoccupied), and calling for some \verb|molecule| larger than H$_2$ in a minimal basis
\begin{lstlisting}[language=Python]
from openfermion.ops import InteractionOperator

core_constant, one_body_integrals, two_body_integrals = (
    molecule.get_active_space_integrals(occupied_indices, active_indices))
active_space_hamiltonian = InteractionOperator(core_constant, 
                                               one_body_integrals, 
                                               two_body_integrals)
\end{lstlisting}
where \verb|active_space_hamiltonian| can now be used to build quantum circuits for the reduced size problem.

\subsection{The InteractionOperator data structure}
As OpenFermion deals primarily with the interactions of physical fermions, especially electrons, the Hamiltonian we have already 
introduced above
\begin{equation}
\label{eq:ferm_ham}
H = h_0 + \sum_{pq} h_{pq} a^\dagger_p a_q + \frac{1}{2} \sum_{pqrs} h_{pqrs} a^\dagger_p a^\dagger_q a_r a_s
\end{equation}
is ubiquitous throughout OpenFermion.

Note that even for fewer than $N=100$ qubits, the coefficients of the terms in the Hamiltonian of \eq{ferm_ham} can require a large amount of memory. Since common Gaussian basis functions lead to the full ${\cal O}(N^4)$ term scaling, instances with less than a hundred qubits can already have tens of millions to hundreds of millions of terms requiring on the order of ten gigabytes to store. Such large Hamiltonians can be expensive to generate (requiring nearly as many integrals as there are terms) so one would often like to be able to save these Hamiltonians. While good for general purpose symbolic manipulation, the FermionOperator data structure is not the most efficient way to store these Hamiltonians, or to manipulate them with efficient numerical linear algebra routines, due to the extra overhead of storing each of the associated operators.

Towards this end we introduce the InteractionOperator data structure. This structure has already been seen in passing in this document in the
first example given.  For example, with a MolecularData object, one may extract the Hamiltonian in the form of an InteractionOperator through
\begin{lstlisting}[language=Python]
h2_hamiltonian = h2_molecule.get_molecular_hamiltonian()
\end{lstlisting}
where the Hamiltonian will be returned as an InteractionOperator.  The InteractionOperator data structure is a class that stores two different matrices and a constant. In the notation of \eq{ferm_ham}, the InteractionOperator stores the constant $h_0$, an $N$ by $N$ array representing $h_{pq}$ and an $N$ by $N$ by $N$ by $N$ array representing $h_{pqrs}$. Note that at present this is not a spatially optimal representation if the goal is specifically to store the integrals of molecular electronic structure systems, but it is a good compromise between space efficiency, simplicity, and ease of performing other numerical operations. The reason it is suboptimal is that there may exist symmetries within the integrals that are not currently utilized.  For example, there is an eight-fold symmetry in the integrals for real basis functions, $h_{pqrs} = h_{sqrp} = h_{prqs} = h_{srqp} = h_{qpsr} = h_{rpsq} = h_{qspr} = h_{rspq}$. Note that for complex basis functions there is only a four-fold symmetry. While we provide methods to iterate over these unique elements, we do not exploit this symmetry in our current implementation for the reason that space efficiency of the InteractionOperator has not yet been a bottleneck in applications. This is consistent with our general design philosophy which is to maintain an active cycle of develop, test, profile, and refine. That is, rather than guess bottlenecks, we analyze and identify the most important ones for problems of interest, and focus time there.

\subsection{The InteractionRDM data structure}
As discussed in the previous section, since fermions are identical particles which interact pairwise, their energy can be determined entirely by reduced density matrices which are polynomial in size. In particular, these energies depend on the one-particle reduced density matrix (1-RDM), denoted here by ${}^{(1)} D$ and the two-particle reduced density matrix (2-RDM), denoted here by ${}^{(2)} D$. The 1-RDM and 2-RDM of a fermionic wavefunction $\ket{\psi}$ are defined through expectation values with one- and two-body local FermionOperators;
\begin{equation}
{}^{(1)} D_{pq} = \bra{\psi} a^{\dagger}_p a_q \ket{\psi}
\quad \quad \quad \quad
{}^{(2)} D_{pqrs} = \bra{\psi} a^{\dagger}_p a_q^\dagger a_r a_s \ket{\psi}.
\label{eq:rdms}
\end{equation}
Thus, the 1-RDM is an $N$ by $N$ matrix and the 2-RDM is an $N$ by $N$ by $N$ by $N$ tensor. Note that the 1-RDM and 2-RDMs may (in general) represent a partial tomography of mixed states, in which case \eq{rdms} should involve traces over that mixed state instead of expectation values with a pure state.  If one has performed a computation at the CISD level of theory, it is possible to extract that density matrix from a molecule using
the following command
\begin{lstlisting}[language=Python]
cisd_two_rdm = h2_molecule.get_molecular_rdm()
\end{lstlisting}

We can see that the energy of $\ket{\psi}$ with a Hamiltonian expressed in the form of \eq{ferm_ham} is given exactly as
\begin{equation}
\avg{E} = h_0 + \sum_{p,q} {}^{(1)} D_{pq} h_{pq} + \frac{1}{2}\sum_{p,q,r,s} {}^{(2)} D_{pqrs} h_{pqrs}
\label{eq:contract_rdm}
\end{equation}
where $h_0$, $h_{pq}$ and $h_{pqrs}$ are the integrals stored by the InteractionOperator data structure.

In OpenFermion, the InteractionRDM class provides an efficient numerical representation of these reduced density matrices. Both InteractionRDM and InteractionOperator inherit from a similar parent class, the PolynomialTensor, reflecting the close parallels between the implementation of these data structures. Due to this parallel, the exact same code which implements integral basis transformations on InteractionOperator also implements integral basis transformations on the InteractionRDM data structure.  Despite their similarities, they represent conceptually distinct concepts,
and in many cases should be treated in a fundamentally different way.  For this reason the implementations are kept distinct.

\subsection{The QuadraticHamiltonian data structure}
  The general electronic structure Hamiltonian \eq{ferm_ham} contains terms that act on up to 4 sites, or
  is quartic in the fermionic creation and annihilation operators. However, in many situations
  we may fruitfully approximate these Hamiltonians by replacing these quartic terms with
  terms that act on at most 2 fermionic sites, or quadratic terms, as in mean-field approximation theory.  
  These Hamiltonians have a number of
  special properties one can exploit for efficient simulation and manipulation of the Hamiltonian, thus
  warranting a special data structure.  We refer to Hamiltonians which
  only contain terms that are quadratic in the fermionic creation and annihilation operators
  as \emph{quadratic Hamiltonians}, and include the general case of non-particle conserving terms as in
  a general Bogoliubov transformation.  Eigenstates of quadratic Hamiltonians can be prepared
  efficiently on both a quantum and classical computer, making them amenable to initial guesses for
  many more challenging problems.

  A general quadratic Hamiltonian takes the form
  \begin{equation}
  \label{eq:quad_ham}
  \sum_{p, q} (M_{pq} - \mu \delta_{pq}) a^\dagger_p a_q
  + \frac12 \sum_{p, q}
  (\Delta_{pq} a^\dagger_p a^\dagger_q + \Delta_{pq}^* a_q a_p)
  + \text{constant},
  \end{equation}
  where $M$ is a Hermitian matrix, $\Delta$ is an antisymmetric matrix,
  $\delta_{pq}$ is the Kronecker delta symbol, and $\mu$ is a chemical
  potential term which we keep separate from $M$ so that we can use it
  to adjust the expectation of the total number of particles.
  In OpenFermion, quadratic Hamiltonians are conveniently represented and manipulated
  using the QuadraticHamiltonian class, which stores $M$, $\Delta$, $\mu$ and
  the constant from \eq{quad_ham}. It is specialized to exploit the properties unique to quadratic Hamiltonians.
  Examples showing the use of this class for simulation are provided later in this document.

\subsection{Some examples justifying data structure design choices}

Here we describe and show a few example illustrating that the above data structures are well designed for efficient calculations that one might do with OpenFermion. These examples are motivated by real use case examples encountered in the authors' own research. We do not provide code examples in all cases here but routines for these calculations exist within OpenFermion.

\subsubsection{FermionOperator example: computation of Trotter error operators}
Suppose that one has a Hamiltonian $H = \sum_{\ell=1}^L H_\ell$ where the $H_\ell$ are single-term FermionOperators. Suppose now that one decides to effect evolution under this Hamiltonian using the second-order Trotter formula, as investigated in works such as \cite{Wecker2014,Hastings2015,Poulin2014,BabbushTrotter,BabbushLow}. A single second-order Trotter step of this Hamiltonian effects evolution under $H + V$ where $V$ is the Trotter error operator which arises due to the fact that the $H_\ell$ do not all commute.  In \cite{Poulin2014} it is shown that a perturbative approximation to the operator $V$ can be expressed as
\begin{equation}
\label{eq:error_op}
V^{(1)} = -\frac{1}{12}\sum_{\alpha \leq \beta} \sum_{\beta} \sum_{\gamma < \beta} \left[H_\alpha\left(1 - \frac{\delta_{\alpha,\beta}}{2}\right), \left[H_{\beta},H_{\gamma}\right]\right].
\end{equation}

Because triangle inequality upper-bounds to this operator provide an estimate of the Trotter error which can be computed in polynomial time, symbolic computation of this operator is crucial for predicting how many Trotter steps one should take in a quantum simulation. However, when using conventional Gaussian basis functions, Hamiltonians of fermionic systems contain ${\cal O}(N^4)$ terms, suggesting that the number of terms in \eq{error_op} could be as high as ${\cal O}(N^{10})$. But in practice, numerics have shown that there is very significant cancellation in these commutators which leads to a number of nontrivial terms that is closer to ${\cal O}(N^4)$ after normal ordering \cite{BabbushTrotter}. If one were to use an array or linked list structure to store FermionOperators then one has two choices. Option (i) is that new terms from the sum are appended to the list and then combined after the sum is completed. Under this strategy the space complexity of the algorithm would scale as ${\cal O}(N^{10})$ and one would likely run out of memory for medium sized systems. Option (ii) is that one normal orders the commutators before adding to the list and then loops through the array or list before adding each term. While this approach has average space complexity of ${\cal O}(N^4)$, the time complexity of this approach would then be ${\cal O}(N^{14})$ as one would need to loop through ${\cal O}(N^4)$ entries in the list after computing each of the ${\cal O}(N^{10})$ terms in the sum. Using our hash table implementation, the average space complexity is still ${\cal O}(N^4)$ but one does not need to loop through all entries at each iteration so the time complexity becomes ${\cal O}(N^{10})$.

Though still quite expensive, one can use Monte Carlo based sampling or distribute this task on a cluster (it is embarrassingly parallel) to compute the Trotter error for medium sized systems. Using the Hamiltonian representations introduced in \cite{BabbushLow}, Hamiltonians have only ${\cal O}(N^2)$ terms, which brings the number of terms in the sum down to ${\cal O}(N^4)$ in the worst case. In that case, computation of $V^{(1)}$ would have time complexity ${\cal O}(N^4)$ using the hash table implementation instead of ${\cal O}(N^6)$ complexity using either a linked-list or an array. The ${\cal O}(N^4)$ complexity enables us to compute $V^{(1)}$ for hundreds of qubits, well into the regime of instances that would be classically intractable to simulate. The task is even more efficient for Hubbard models which have ${\cal O}(N)$ terms.

\subsubsection{FermionOperator example: symbolic Fourier transformation}

A secondary goal for OpenFermion is that the library can be used to as a tool for the symbolic manipulation of fermionic Hamiltonians in order to analyze and develop new simulation algorithms and Hamiltonian representations. To give a concrete example of this, in the recent paper \cite{BabbushLow}, authors were able to demonstrate Trotter steps of the electronic structure Hamiltonian with significantly reduced complexity: ${\cal O}(N)$ depth as opposed to ${\cal O}(N^4)$ depth. A critical component of that improvement was to represent the Hamiltonian using basis functions which are a discrete Fourier transform of the plane wave basis (the plane wave dual basis) \cite{BabbushLow}. The appendices of that paper begin by showing the Hamiltonian in the plane wave basis:
 \begin{equation}
 H = \frac{1}{2} \sum_{p, \sigma} k_p^2 \, c_{p,\sigma}^\dagger c_{p,\sigma} - \frac{4 \pi}{\Omega} \sum_{\substack{p \neq q \\ j,\sigma}} \left(\zeta_j \frac{e^{i \, k_{p-q} \cdot R_j}}{k_{q-p}^2}\right) c^\dagger_{p, \sigma} c_{q, \sigma} + \frac{2 \pi}{\Omega} \sum_{\substack{(p, \sigma) \neq (q, \sigma') \\ \nu \neq 0}} \frac{c^\dagger_{p,\sigma} c_{q,\sigma'}^\dagger c_{q + \nu,\sigma'} c_{p - \nu,\sigma}}{k_\nu^2}.
 \label{eq:pw_ham}
 \end{equation}
To obtain the Hamiltonian in the plane wave dual basis, one applies the Fourier transform of the mode operators,
 \begin{equation}
\label{eq:ladder_def}
c^\dagger_\nu =  \sqrt{\frac{1}{N}}  \sum_{p} a^\dagger_{p} e^{- i \,k_\nu \cdot r_p}
\quad \quad \quad \quad
c_\nu =  \sqrt{\frac{1}{N}}  \sum_{p} a_{p} e^{i \, k_\nu \cdot r_p}.
\end{equation}
Using OpenFermion one can easily generate the plane wave Hamiltonian (either manually or by using the plane wave module) and then apply the discrete Fourier transform of the mode operators (either manually or by using the discrete Fourier transform module) to verify the correct form of the plane wave dual Hamiltonian shown in \cite{BabbushLow},
\begin{align}
\label{eq:pwd_ham}
H & = \frac{1}{2\, N} \sum_{\nu, p, q, \sigma} k_\nu^2 \cos \left[k_\nu \cdot r_{q - p} \right] a^\dagger_{p, \sigma} a_{q,\sigma}
- \frac{4 \pi}{\Omega} \sum_{\substack{p,\sigma \\ j, \nu\neq 0}} \frac{\zeta_j \, e^{i \, k_{\nu} \cdot \left(R_j - r_{p}\right)} }{k_\nu^2} n_{p, \sigma} + \frac{2 \pi}{\Omega } \sum_{\substack{(p, \sigma) \neq (q, \sigma') \\ \nu \neq 0}} \frac{\cos \left[k_\nu \cdot r_{p-q}\right]}{k_\nu^2} \, n_{p, \sigma} n_{q, \sigma'}.
\end{align}.

While \eq{pwd_ham} turns out to have a very compact representation, it requires a careful derivation to show that application of \eq{ladder_def} to \eq{pw_ham} leads to \eq{pwd_ham}. However, this task is trivial for OpenFermion since the Fourier transform can be applied symbolically and the output FermionOperator can be simplified automatically using a normal-ordering routine. This example demonstrates the utility of OpenFermion for verifying analytic calculations.

\subsubsection{InteractionOperator example: fast orbital basis transformations}

An example of an important numerical operation which is particularly efficient in this representation is a rotation of the molecular orbitals $\phi_p\left(r\right)$. This unitary basis transformation takes the form
\begin{equation}
\tilde{\phi}_p\left(r\right) = \sum_{q=1}^N \phi_q \left(r\right) U_{pq} \quad \quad
U = e^{- \kappa} \quad  \quad \kappa = - \kappa^\dagger = \sum_{p,q} \kappa_{pq} a^\dagger_p a_q.
\label{eq:rotation}
\end{equation}
For $\kappa$ and $U$, the quantities $\kappa_{pq}$ and $U_{pq}$ respectively correspond to the matrix elements of these $N$ by $N$ matrices. We see then that the ${\cal O}(N^2)$ elements of the matrix $\kappa$ define a unitary transformation on all orbitals. Often, one would like to apply this transformation to Hamiltonian in the orbital basis defined by $\{\phi_p(r)\}$ in order to obtain a new Hamiltonian in the orbital basis defined by $\{\tilde{\phi}_p(r)\}$. Since computation of the integrals is extremely expensive, the goal is usually to apply this transformation directly to the Hamiltonian operator.

When specialized to real, orthogonal rotations $U$ (which is often the case in molecular electronic structure), the most straightforward expression of this integral transformation for the two-electron integrals is
\begin{equation}
\tilde{h}_{pqrs} = \sum_{\mu,\nu,\lambda,\sigma} U_{p\mu} U_{q\nu} h_{\mu \nu \lambda \sigma} U_{r\lambda} U_{s\sigma}.
\end{equation}
Since there are ${\cal O}(N^4)$ integrals, the entire transformation of the Hamiltonian would take time ${\cal O}(N^8)$, which is extremely onerous. A more efficient approach is to rearrange this expression to obtain the following,
\begin{equation}
\tilde{h}_{pqrs} = \sum_{\mu} U_{p\mu} \left[\sum_{\nu} U_{q\nu} \left[ \sum_{\lambda} U_{r\lambda} \left[ \sum_{\sigma} U_{s\sigma} h_{\mu \nu \lambda \sigma} \right]\right] \right]
\end{equation}
which can be evaluated for each term at cost ${\cal O}(N)$ since each summation can be carried out independently. This brings the total cost of the integral transformation to ${\cal O}(N^5)$. While such a transformation would be extremely tedious using the FermionOperator representation, this approach is implemented for InteractionOperators in OpenFermion using the \verb|einsum| function for numerical linear algebra from \verb|numpy|.  

This functionality is readily accessible within OpenFermion for basis rotations.  For example, the one may rotate the basis of the molecular hamiltonian of H$_2$ to a new basis with some orthogonal matrix $U=\exp(-\kappa)$ of appropriate dimension as 
\begin{lstlisting}[language=Python]
from numpy import array, eye, kron

U = kron(array([[0, 1], [1, 0]]), eye(2))
h2_hamiltonian = h2_molecule.get_molecular_hamiltonian()
h2_hamiltonian.rotate_basis(U)
\end{lstlisting}
We now provide an example of where such fast basis transformations may be useful. Previous work has shown that the number of measurements required for a variational quantum algorithm to estimate the energy of a Hamiltonian such as \eq{ferm_ham} to precision $\epsilon$ scales as
\begin{equation}
M = {\cal O}\left(\left[\frac{1}{\epsilon} \sum_{p,q,r,s} \left| h_{pqrs} \right | \right]^2\right).
\end{equation}
Since the $h_{pqrs}$ are determined by the orbital basis, one can alter this bound by rotating the orbitals under angles organized into the $\kappa$ matrix of \eq{rotation}. Thus, one may want to perform an optimization over these angles in order to minimize $M$. This task would be quite unwieldy or perhaps nearly impossible using the FermionOperator data structure but is viable using the InteractionOperator data structure with fast integral transformations.

\subsubsection{QuadraticHamiltonian example: preparing fermionic Gaussian states}
  As mentioned above, eigenstates of quadratic Hamiltonians, \eq{quad_ham}, can be
  prepared efficiently on a quantum computer, and OpenFermion includes functionality
  for compiling quantum circuits to prepare these states.
  Eigenstates of general quadratic Hamiltonians with both particle-conserving and non-particle conserving terms
  are also known as fermionic Gaussian states. A key step in the preparation of
  a fermionic Gaussian state is the computation of a basis transformation which
  puts the Hamiltonian of \eq{quad_ham} into the form
  \begin{equation}
    \label{eq:quad_ham_free}
    \sum_p \varepsilon_p b^\dagger_p b_p + \text{constant},
  \end{equation}
  where the $b^\dagger_p$ and $b_p$ are a new set of fermionic creation and
  annihilation operators that also obey the canonical fermionic anticommutation relations.
  In OpenFermion, this basis transformation is computed efficiently with
  matrix manipulations by exploiting the special properties of quadratic Hamiltonians
  that allow one to work with only the matrices $M$ and $\Delta$ from \eq{quad_ham}
  stored by the QuadraticHamiltonian class, using simple numerical linear algebra
  routines included in SciPy.
  The following code constructs a mean-field Hamiltonian of the \emph{d}-wave  model of superconductivity and then obtains a circuit that prepares its ground state 
  (along with a description of the starting state to which the circuit should be applied):

\begin{lstlisting}[language=Python]
from openfermion.hamiltonians import mean_field_dwave
from openfermion.transforms import get_quadratic_hamiltonian
from openfermion.utils import gaussian_state_preparation_circuit

x_dimension = 2
y_dimension = 2
tunneling = 2.
sc_gap = 2.
periodic = True

mean_field_model = mean_field_dwave(x_dimension, y_dimension, tunneling, sc_gap, periodic)
quadratic_hamiltonian = get_quadratic_hamiltonian(mean_field_model)
circuit_description, start_orbitals = gaussian_state_preparation_circuit(quadratic_hamiltonian)
\end{lstlisting}
The circuit description follows the procedure explained in~\cite{Jiang2017}. One can also obtain the ground energy and ground state numerically with the following code:
\begin{lstlisting}[language=Python]
from openfermion.utils import jw_get_gaussian_state
ground_energy, ground_state = jw_get_gaussian_state(quadratic_hamiltonian)
\end{lstlisting}

\section{Models and utilities}
OpenFermion has a number of capabilities that assist with the creation and manipulation of fermionic and related Hamiltonians.  While they are too numerous and growing too quickly to list in their entirety, here we will briefly describe a few of these examples in an attempt to ground what one would expect to find within OpenFermion.

One broad class of functions found within OpenFermion is the generation of fermionic and related Hamiltonians.  While the MolecularData structure discussed above is one example for molecular systems, we also include utilities for a number of model systems of interest as well.  For example, there are currently supported routines for generating Hamiltonians of the Hubbard model, the homogeneous electron gas (jellium), general plane wave discretizations, and $d$-wave models of superconductivity.

To give a concrete example, if one wished to study the homogeneous electron gas in 2D, which is of interest when studying the fractional quantum hall effect, then one could use OpenFermion to initialize the model as follows
\begin{lstlisting}[language=Python]
from openfermion.hamiltonians import jellium_model
from openfermion.utils import Grid

jellium_hamiltonian = jellium_model(Grid(dimensions=2, 
                                         length=10, 
                                         scale=1.0))
\end{lstlisting}
where jellium\_hamiltonian will be a fermionic operator representing the spinful homogeneous electron gas discretized into a $10 \times 10$ grid of plane waves in two dimensions.  One is then free to transform this Hamiltonian into qubit operators and use it in the algorithm of choice.

Similarly, one may use the utilities provided to prepare a Fermi-Hubbard model for study.  For example
\begin{lstlisting}[language=Python]
from openfermion.hamiltonians import fermi_hubbard

t = 1.0
U = 4.0

hubbard_hamiltonian = fermi_hubbard(x_dimension=10, y_dimension=10, 
                                    tunneling=t, coulomb=U,
                                    chemical_potential=0.0, periodic=True)
\end{lstlisting}
creates a Fermi-Hubbard Hamiltonian on a square lattice in 2D that is $10 \times 10$ sites with periodic boundary conditions.

Besides Hamiltonian generation, OpenFermion also includes methods for outputting Trotter-Suzuki decompositions of arbitrary operators, providing the corresponding quantum circuit in QASM format \cite{Cross2017}. These methods were included in order to simplify porting the resulting quantum circuits to other simulation packages, such as LIQUi$|>$ ~\cite{liq1},  qTorch \cite{Fried:2017}, Project-Q \cite{Steiger2016}, and qHipster \cite{qhipster}. For instance, the function \texttt{pauli\_exp\_to\_qasm} takes a list of QubitOperators and an optional evolution time as input, and outputs the QASM specification as a string. As an example,
\begin{lstlisting}[language=Python,upquote=true]
from openfermion.ops import QubitOperator
from openfermion.utils import pauli_exp_to_qasm

for line in pauli_exp_to_qasm([QubitOperator('X0 Z1 Y3', 0.5), QubitOperator('Z3 Z4', 0.6)]):
    print(line)
\end{lstlisting}
outputs a QASM specification for a circuit corresponding to $U = e^{-i 0.5 X_0 Z_1 Y3} e^{-i 0.6 Z_3 Z_4}$, which in this case is given by
\begin{lstlisting}
H 0
Rx 1.5707963267948966 3
CNOT 0 1
CNOT 1 3
Rz 0.5 3
CNOT 1 3
CNOT 0 1
H 0
Rx -1.5707963267948966 3
CNOT 3 4
Rz 0.6 4
CNOT 3 4
\end{lstlisting}

OpenFermion additionally supports a number of other tools including the ability to evaluate the Trotter error operator and construct circuits for preparing arbitrary Slater determinants on a quantum computer using the linear depth procedure described in \cite{Jiang2017}.  Some support for numerical simulation is included for testing purposes, but the heavy lifting in this area is delegated to plugins specialized for these applications. In the future, we imagine these utilities will expand to include more Hamiltonians and specializations that add in the creation of simulation circuits for fermionic systems.

\section{Open Source Management and Project Philosophy}

OpenFermion is designed to be a tool for both its developers and the community at large.  By maintaining an open-source and framework
independent library, we believe it provides a useful tool for developers in industry, academia, and research institutions alike.
Moreover, it is our hope that these developers will, in turn, contribute code they found to be useful in interacting with OpenFermion
to the benefit of the field as a whole. Here we outline some of our philosophy, as well as the ways in which we ensure that OpenFermion
remains a high quality package, even as many different developers from different institutions contribute.

\subsection{Style and testing}
The OpenFermion code is written primarily in Python with optional C++ backends being introduced as higher performance is desired.
Stylistically, this code follows the PEP8 guidelines for Python and demands descriptive names as well as extensive documentation
for all functions.  This enhances readability and allows developers to more reliably build off of contributed code.  

At present, the source code is managed through GitHub where the standard pull request system is used for making contributions.
When a pull request is made, it must be reviewed by at least one member of the OpenFermion team who has experience with the library
and is able to comment on both the style and integration prospects with the library.  As contributors become more familiar with the
code, they may become approved reviewers themselves to enhance their contribution to the process. All contributors are welcome to assist with reviews but reviews from new contributors will not automatically enable merging into the master branch.

Tests in the code are written in the python unittest framework and all code is required to be both Python 2 and Python 3 compliant so that it continues
to be as useful as possible for all users in the future.  The tests are run automatically when a pull request is made through
the Travis continuous integration (CI) framework, to minimize the chance that code that will break crucial functionality is accidentally
merged into the code base, and ensure a smooth development and execution experience.

\subsection{Distribution}
Several options for code distribution are available to obtain OpenFermion.  The code may be pulled and used directly from the 
GitHub repository if desired (which one can find at \href{http://www.openfermion.org}{\texttt{www.openfermion.org}}), and the requirements may be manually fulfilled by the user.  This option offers maximum control and access
to bleeding edge code and developments, but minimal convenience. A full service option is offered through the Python Package Index (PyPI)
so that installation for most users can be as simple as 
\begin{lstlisting}
python -m pip install openfermion
\end{lstlisting}
A middle ground between these options which is popular with many of the lead developers of this project is to pull the latest code directly from GitHub but to install with pip using the development install command in the OpenFermion directory:
\begin{lstlisting}
python -m pip install -e .
\end{lstlisting}
In the future, a version of the code may be supported for other distribution platforms as well.

The OpenFermion plugins (and the packages on which they rely) need to be installed independently using a similar procedure. One can install OpenFermion-Psi4 using pip under the name ``openfermionpsi4'', OpenFermion-Cirq under the name ``openfermioncirq'' and OpenFermion-PySCF under the name ``openfermionpyscf". These plugins can also be installed directly from GitHub (we link to repositories from the main OpenFermion page). 

In addition to the traditional installation models of either installing from the PyPI registry, or downloading the source from GitHub and installing, the project also supports a Docker container.  Docker containers offer a compact virtualization environment that is portable between all systems where Docker is supported.  This is a convenient option for both first time users, and those who want to deploy OpenFermion to non-standard architectures (or on Windows).  Moreover, it offers easy access to some of the electronic structure packages that our plugins are inter-operable with, which allows users convenient access to the full tool chain.  At present, the Dockerfile is hosted on the repository, which can be used to build a Docker image as detailed in full in the Readme of the Docker folder.

\section*{Closing Remarks}
The rapid development of quantum hardware represents an impetus for the equally rapid development of quantum applications.  Development of these applications represents a unique challenge, often requiring the expertise or domain knowledge from both the application area and quantum algorithms.  OpenFermion is a bridge between the world of quantum computing and materials simulation, which we believe will act as a catalyst for development in this critical area.  Only with such software tools can we start to explore the explicit costs and advantages of new algorithms, and push forward with practical advancements.  It is our hope that OpenFermion not only leads to developments in the field of quantum computation for quantum chemistry and materials, but also sparks the development of similar packages for other application areas.

\section*{Acknowledgements}

The authors thank Hartmut Neven for encouraging the initiation of this project at Google as well as Al\'{a}n Aspuru-Guzik, Carlo Beenakker, Yaoyun Shi, Matthias Troyer, Stephanie Wehner and James Whitfield for supporting graduate student and postdoc developers who contributed code to OpenFermion. 
I.\ D.\ K.\ acknowledges partial support from the National Sciences and Engineering Research Council of Canada. K.\ J.\ S.\ acknowledges support from NSF Grant No. 1717523. T.\ H.\ and D.\ S.\ S.\ have been supported by the Swiss National Science Foundation through the National Competence Center in Research QSIT. S.\ S.\ is supported by the DOE Computational Science Graduate Fellowship under grant number DE-FG02-97ER25308. MS is supported by the Netherlands Organization for Scientific Research (NWO/OCW) and an ERC Synergy Grant.

\bibliographystyle{apsrev4-1}
\bibliography{references,Mendeley,library}

\end{document}